\documentstyle[12pt,epsfig]{article}
\newcommand{\be}{\begin{equation}}
\newcommand{\ee}{\end{equation}}
\newcommand{\ba}{\begin{eqnarray}}
\newcommand{\ea}{\end{eqnarray}}

\newcommand{\nsigma}{\mbox{\boldmath $\sigma$}}
\newcommand{\nkappa}{\mbox{\boldmath $\kappa$}}

\newcommand{\neta}{\mbox{\boldmath $\eta$}}

\newcommand{\nl}{{\bf      l}}

\newcommand{\nk}{{\bf      k}}
\newcommand{\np}{{\bf      p}}
\newcommand{\nq}{{\bf      q}}
\newcommand{\nr}{{\bf      r}}

\newcommand{\nv}{{\bf      v}}

\newcommand{\nJ}{{\bf      J}}

\newcommand{\nY}{{\bf      Y}}

\newcommand{\hr}{{\bf \hat{r}}}

\newcommand{\ninej}[9]{\left\{
                         \begin{array}{ccc}
                           \textstyle #1 &
                           \textstyle #2 &
                           \textstyle #3 \\
                           \textstyle #4 &
                           \textstyle #5 &
                           \textstyle #6 \\
                           \textstyle #7 &
                           \textstyle #8 &
                           \textstyle #9
                          \end{array}
                       \right\}}

\newcommand{\sixj}[6]{\left\{
                         \begin{array}{ccc}
                           \textstyle #1 &
                           \textstyle #2 &
                           \textstyle #3 \\
                           \textstyle #4 &
                           \textstyle #5 &
                           \textstyle #6
                          \end{array}
                       \right\}}

\newcommand{\threej}[6]{\left(
                         \begin{array}{ccc}
                           \textstyle #1 &
                           \textstyle #2 &
                           \textstyle #3 \\
                           \textstyle #4 &
                           \textstyle #5 &
                           \textstyle #6
                          \end{array}
                       \right)}

\begin{document}
\begin{titlepage}
\mbox{}
\vspace*{2.5\fill}
{\Large\bf
\begin{center}
Semi-relativistic description of quasielastic
neutrino reactions and superscaling
in a continuum shell model
\end{center}
}
\vspace{1\fill}
\begin{center}
{\large
J.E. Amaro$^{1}$,
M.B. Barbaro$^{2}$,
J.A. Caballero$^{3}$,
T.W. Donnelly$^4$,
C. Maieron$^5$
}
\end{center}
\begin{small}
\begin{center}
$^{1}${\em
          Departamento de F\'{\i}sica Moderna,
          Universidad de Granada,
          Granada 18071, Spain}  \\[2mm]
$^{2}${\em
          Dipartimento di Fisica Teorica,
          Universit\`a di Torino and
          INFN, Sezione di Torino \\
          Via P. Giuria 1, 10125 Torino, Italy} \\[2mm]
$^{3}$ {\em
          Departamento de F\'\i sica At\'omica, Molecular y Nuclear \\
          Universidad de Sevilla, Apdo. 1065, E-41080 Sevilla, Spain
}\\[2mm]
$^4$ {\em Center for Theoretical Physics,
          Laboratory for Nuclear Science
          and Department of Physics,
          Massachusetts Institute of Technology,
          Cambridge, MA 02139, USA
}\\[2mm]
$^5$ {\em INFN, Sezione di Catania,
              Via Santa Sofia 64,
              95123 Catania, Italy
}
\end{center}
\end{small}

\kern 1. cm \hrule \kern 3mm

\begin{small}
\noindent
{\bf Abstract}
\vspace{3mm}

\noindent
The so-called semi-relativistic expansion of the weak charged current in powers
of the initial nucleon momentum is performed to describe charge-changing,
quasielastic neutrino reactions $(\nu_\mu,\mu^-)$ at intermediate
energies. The quality of the expansion is tested by comparing with the
relativistic Fermi gas model using several choices of kinematics of interest for ongoing
neutrino oscillation experiments.  The new current is then implemented in a
continuum shell model together with relativistic kinematics to investigate
the scaling properties of $(e,e')$ and $(\nu_\mu,\mu^-)$ cross sections.

\kern 2mm

\noindent
{\em PACS:}\
25.30.Pt;  
25.30.Fj,  
24.10.Jv,  

\noindent
{\em Keywords:}\ Nuclear reactions; Neutrino scattering;
Quasielastic electron scattering; Scaling;
Relativistic models.

\end{small}

\kern 2mm \hrule \kern 1cm
\end{titlepage}


\section{Introduction}


The importance of neutrino-induced reactions in nuclei has been
stressed in connection with the neutrino oscillation experiments
performed by the KARMEN and LSND collaborations
\cite{Kra92}--\cite{Aue02}.  In experiments of this type
relatively low $\nu_e$ or $\nu_\mu$ energies are involved (at most
a few hundreds of MeV) and so the nuclear excitations involved can
be described by standard non-relativistic models of the reaction
including the relevant machinery (RPA correlations, large-basis
shell models, $\Delta$-hole excitations, final-state interactions,
{\it etc.}) for this energy regime, where in particular giant
resonances may play an important role \cite{Ko94}--\cite{Krm04}.

However, when passing to the ongoing and next generations of
neutrino experiments, MiniBooNE, K2K/T2K, MINOS, NO$\nu$A, and
MINER$\nu$A \cite{Fuk98}--\cite{Dra04}, the neutrino beam energies
increase to the GeV level, and typically large energies and
momenta are transferred to the nucleus. For these kinematics
relativity is important and non-relativistic models of the
reaction such as those listed above are bound to fail unless the
relevant relativistic ingredients are included.

First of all the use of relativistic
kinematics is required, and must be implemented in the model. From
quasielastic (QE) electron scattering studies we know that a good
approximation to the correct kinematics consists in the substitution
\begin{equation}
\lambda\longrightarrow \lambda(1+\lambda)   \label{kinematics}
\end{equation}
with $\lambda=\omega/2m_N$, where $\omega$ is the energy transfer and
$m_N$ the nucleon mass. This substitution can easily be performed in
all places in the calculation --- except in the nucleon form factors,
where the correct value of momentum transfer $Q^\mu=(\omega,\nq)$ must be
used.  Second, a good approximation to the current matrix elements is
required.  The current is traditionally obtained from the fully
relativistic one by some expansion procedure, usually through the
Foldy-Wouthuysen transformation, which is valid for small momenta
compared with the nucleon mass.  Within this procedure relativistic
effects were studied for neutrino energies up to 300 MeV in
\cite{Kuramoto:1989tk}, showing that a good agreement is obtained
between fully relativistic and non-relativistic Fermi gas
calculations, when terms up to order $(q/m_N)^3$ are included in the
latter. Similar results were obtained at larger energies, for the
scattering of atmospheric neutrinos from oxygen \cite{Engel:1993nn},
where, however, it must be noted that, due to the steep decrease of
the atmospheric neutrino spectrum at large energies
\cite{Barr:2004br}, the most important contributions to the process
correspond to relatively small energy transfers.  Obviously the
Foldy-Wouthuysen expansion is not applicable for the high values of
the momentum transfer $q$ which are of interest for the new
experiments, {\it i.e.,} for values around 1 GeV/c. In this case a
different expansion procedure must be performed in which $q$ and
$\omega$ can be arbitrarily large.

The first goal of this paper is thus to develop an approximation
to the nuclear charged current (CC) that accounts for specific relativistic
effects which are relevant to intermediate energy quasielastic neutrino
reactions.  The CC is obtained here through an expansion that only
requires the momentum of the initial nucleon to be small, while it
treats exactly the dependence on $(\omega,\nq)$.  Moreover the
resulting semi-relativistic (SR) current is simple enough to be
easily implemented in already existing non-relativistic models of
$(\nu_l,l^-)$ and $(\overline{\nu}_l,l^+)$ reactions, where $l=e$
or $\mu$.  It is presented here as an extension of the
electromagnetic and weak-neutral current expansion originally
derived in \cite{Ama96a} and that in recent years has been
widely tested and applied in several collaborations to describe
a wide variety of inclusive and exclusive electron
scattering observables for intermediate energies and excitations
in the vicinity of the QE peak
\cite{Ama96b}--\cite{Ama05a}.  Extensions of the SR expansion to
meson-exchange currents have also been developed
\cite{Ama98b,Ama03b}, and a detailed description of their
application to two-body currents can be found in a recent review
article \cite{Ama02b}.

In this paper we apply the SR model to neutrino-induced reactions for the
first time.  Apart from the different isospin dependence, the SR expansion
coincides essentially with that of the weak neutral current performed in
\cite{Ama96a}.  The only difference is that in the present reaction we have to
include the time component of the axial-vector current. That component was not
considered in \cite{Ama96a} because it does not
contribute significantly to parity-violating electron scattering.

We check the quality of the SR expansion in the context of the
Relativistic Fermi Gas (RFG) model.  This is a very convenient
model for our purposes, since it is fully relativistic and simple
enough to be solved exactly.  Also this model is capable of
getting the basic size and shape of the QE $(e,e')$ cross section.
Thus we shall show that, starting from the non-relativistic Fermi
gas, performing the replacement in Eq.~(\ref{kinematics}) in the
kinematics and implementing the new SR charged current, we
reproduce basically the RFG results for $(\nu_\mu,\mu^-)$
reactions. This ``relativizing '' procedure can easily be
extrapolated to more sophisticated finite nuclei models of the
reaction.  The result will at least reproduce appropriately the
allowed kinematical region and the relevant relativistic content
of the current operator. Of course features related to the
relativistic aspects of the dynamics cannot be accounted for by
our procedure. 
In context we note that studies of relativistic nuclear
dynamics in charged and neutral-current neutrino-nucleus
QE scattering have already been presented in some
previous work~\cite{Alb97,Alb98,Alb99,Chiara03,Meucci04}. In these
studies a basic focus was to analyze the effects introduced by
various descriptions of the final nucleon relativistic states upon
the integrated cross sections.

We illustrate the relativizing procedure by applying it to the
continuum shell model (CSM), {\it i.e.,} nucleons in a mean field taken here
to be a Woods-Saxon potential. We use the same (real)
potential for initial and final states in order to maintain
orthogonality between nuclear states.  In this way we extend the SR
shell model of \cite{Ama96a} to neutrino reactions at the QE
peak. Since the use of relativistic kinematics is potentially
equivalent to solving a Klein-Gordon equation, the present SR
continuum shell model includes some aspects of a relativistic mean
field.

With such a model we are in a position to fulfill the second goal of
this paper, which is to investigate superscaling properties of both
$(e,e')$ and $(\nu_\mu,\mu^-)$ inclusive cross sections at the QE peak
for intermediate energies, namely the degree to which one finds that
the reduced cross sections are independent of the momentum transfer
(scaling of the first kind) or the nuclear species (scaling of the
second kind) or both (superscaling).  Exhaustive analyses of the
$(e,e')$ world data and explorations of various aspects of their
scaling properties have been performed in \cite{Alb88}--\cite{Bar04}.
In particular, in recent work \cite{Ama05b} the approach has been extended 
to the $\Delta$ peak, allowing one to construct a
semi-empirical model based on scaling which is very successful in
describing the experimental $(e,e')$ cross section up to the $\Delta$
peak for high energies. This model allowed us in \cite{Ama05b} to
generate predictions for the $(\nu_\mu,\mu^-)$ cross section, under
the reasonable hypothesis that it presents, for high energies, scaling
properties similar to those of the $(e,e')$ cross section.  This
hypothesis is true by construction in the RFG and can be also
demonstrated at least for the conventional Plane Wave Impulse
Approximation (PWIA) in the high $q$ limit~\footnote{However, this is
broken to some degree by the Relativistic PWIA (RPWIA) where
factorization no longer obtains.}.  However, when distortion of the
ejected nucleon is present, a general proof of scaling cannot be
provided, while for lower energies it is clear that the axial-vector
and vector matrix elements are renormalized differently in the nuclear
medium due to RPA correlations, and this can obviously modify the
scaling properties of the neutrino cross section.  For higher energies
one expects that nuclear effects such as from RPA correlations are
less important than at low energies. In this work we use the SR shell
model to investigate the degree of violation of the scaling hypothesis
for the QE peak region within the Distorted Wave Impulse Approximation
(DWIA), where the distortion of the ejected nucleon is described with
a real potential.  We perform this study in two steps. First we focus
on the $(e,e')$ cross section and study the scaling properties of the
separate response functions.  Once the superscaling has been verified
for electromagnetic processes, we are able to reconstruct the
$(\nu_\mu,\mu^-)$ cross section from the $(e,e')$ one using the
scaling hypothesis and compare it with the one computed directly using
the SR shell model.  Thus, at least within the context of the SR shell
model discussed in the present work, following this procedure we shall
be able to check the consistency of our approach and quantify the
degree to which scale-breaking effects are expected to enter. As we
shall see below, there appears to be very little impact from this
source of scale-breaking on the scaling approach used in \cite{Ama05b}
to predict neutrino-induced cross sections.

The paper is organized in the following way.  In Sec.~2 we begin with
a brief review of the general formalism for neutrino scattering, and
present the expansion of the CC. We particularize the formalism for
the shell model, and introduce the general multipole expansion of the
responses, with some details on the derivation of the Coulomb
multipoles of the axial-vector current placed in Appendix~A.  We also
provide the expressions for the factorized PWIA in Appendix~B.  In
Sect.~3 we present results for the $(\nu_\mu,\mu^-)$ reaction for
several choices of kinematics and several nuclei of interest.  We
first perform an analysis of the quality of the relativizing procedure
by comparison with the RFG for the relevant kinematics. The quality of
the various components of the SR current can be checked separately by
examining the individual response functions that contribute to the
process.  
We then focus on the SR shell model and perform the
scaling analysis of the longitudinal and transverse electromagnetic
responses as functions of the momentum transfer for various nuclei.
We apply the scaling hypothesis to reconstruct the $(\nu_\mu,\mu^-)$
cross section starting from the electromagnetic scaling function and
compare with the shell model results.  Finally, in Sec. 4 we present
our conclusions.

\section{Formalism}

In this section we briefly present the basic formalism for neutrino-induced
reactions.  Some of the previous approaches to the general formalism for these
reactions can be found in \cite{Alb97}, \cite{Don79}--\cite{Barbaro:1996vd}.

\subsection{Charge-changing neutrino cross section}

Here we focus on the particular case of $(\nu_\mu,\mu^-)$, while the cases of
anti-neutrinos or of other lepton species can be easily obtained with obvious
changes.  The four-momenta of the incident neutrino and detected muon are
$k^\mu=(\epsilon,\nk)$ and $k'{}^\mu=(\epsilon',\nk')$, respectively.  The
four-momentum transfer is $Q^\mu=k^\mu-k'{}^\mu=(\omega,\nq)$.  We use a
coordinate system with the $z$-axis pointing along $\nq$ and the $x$-axis
along the transverse component of the incident momentum, {\it i.e.,}
$\nk^\perp=\nk-\frac{1}{q^2}(\nk\cdot\nq)\nq$.  We follow the formalism of
\cite{Ama05b} where the cross section is written as
\begin{equation}
\frac{d\sigma}{d\Omega'd\epsilon'}=\sigma_0{\cal F}_+^2 \ ,
\end{equation}
with
\begin{equation}
\sigma_0=
\frac{G^2\cos^2\theta_c}{2\pi^2}
k'\epsilon'\cos^2\frac{\tilde{\theta}}{2}\ .
\end{equation}
Here $G=1.166\times 10^{-11}\quad\rm MeV^{-2}$ is the Fermi constant,
$\theta_c$ is the Cabibbo angle, $\cos\theta_c=0.975$, and the angle
$\tilde\theta$ is defined as
\begin{equation}
\tan^2\frac{\tilde{\theta}}{2}=\frac{|Q^2|}{(\epsilon+\epsilon')^2-q^2}\ ,
\end{equation}
with $Q^2=\omega^2-q^2 < 0$.
The nuclear structure information is contained in ${\cal F}_+^2$, defined by
\begin{equation}
{\cal F}_+^2=
\widehat{V}_{CC} R_{CC}+
2\widehat{V}_{CL} R_{CL}+
\widehat{V}_{LL} R_{LL}+
\widehat{V}_{T} R_{T}+
2\widehat{V}_{T'} R_{T'}\ ,
\end{equation}
where the kinematical factors $\widehat{V}_K$ coming from the leptonic tensor
are defined by
\begin{eqnarray}
\widehat{V}_{CC}
&=&
1-\delta^2\tan^2\frac{\tilde{\theta}}{2}
\label{vcc}\\
\widehat{V}_{CL}
&=&
\frac{\omega}{q}+\frac{\delta^2}{\rho'}\tan^2\frac{\tilde{\theta}}{2}
\\
\widehat{V}_{LL}
&=&
\frac{\omega^2}{q^2}+
\left(1+\frac{2\omega}{q\rho'}+\rho\delta^2\right)\delta^2
\tan^2\frac{\tilde{\theta}}{2}
\\
\widehat{V}_{T}
&=&
\tan^2\frac{\tilde{\theta}}{2}+\frac{\rho}{2}-
\frac{\delta^2}{\rho'}
\left(\frac{\omega}{q}+\frac12\rho\rho'\delta^2\right)
\tan^2\frac{\tilde{\theta}}{2}
\\
\widehat{V}_{T'}
&=&
\frac{1}{\rho'}
\left(1-\frac{\omega\rho'}{q}\delta^2\right)
\tan^2\frac{\tilde{\theta}}{2}\ .
\label{vtp}
\end{eqnarray}
In Eqs.~(\ref{vcc}--\ref{vtp}), following \cite{Ama05b},
we have defined
\begin{eqnarray}
\delta &=& \frac{m'}{\sqrt{|Q^2|}}\\
\rho &=& \frac{|Q^2|}{q^2}\\
\rho' &=& \frac{q}{\epsilon+\epsilon'}\ .
\end{eqnarray}
Note that the only dependence on the muon mass $m'$ is contained in the
$\delta$ coefficient.

Finally the weak response functions are given by
\begin{eqnarray}
R_{CC} &=&W^{00} \label{RC}\\
R_{CL} &=& -\frac12\left( W^{03}+W^{30} \right)\\
R_{LL}  &=& W^{33}  \\
R_T &=&  W^{11}+W^{22} \\
R_{T'} &=& -\frac{i}{2}\left( W^{12}-W^{21} \right) \label{RTP}
\end{eqnarray}
in terms of the inclusive hadronic tensor \cite{Don79}:
\begin{equation} \label{hadronic}
W^{\mu\nu}(q,\omega)= \overline{\sum_{fi}}\delta(E_f-E_i-\omega)
\langle f | {J}^\mu(Q)|i\rangle^*
\langle f | {J}^\nu(Q)|i\rangle\ .
\end{equation}
In Eq.~(\ref{hadronic}), $J^\mu(Q)$ is the hadronic CC current
operator, to be specified below, and a sum over final states and
an average over initial spin is assumed.

\subsection{Semi-relativistic charge-changing current}

We begin with the basic relativistic charged weak current of the
nucleon, $j^\mu=j^\mu_V-j^\mu_A$. 
In this work we employ only the Standard Model of electroweak interactions
at tree level and thus, for example, do not include radiative corrections or
contributions from second-class currents (see \cite{Don79}). 
We use the conventions of
\cite{Bjo64}. The vector and axial-vector currents are given by
\begin{eqnarray}
j^\mu_V(\np',\np)
&=&
\overline{u}(\np')
\left[ 2F_1^V \gamma^\mu + i\frac{F_2^V}{m_N}\sigma^{\mu\nu}Q_\nu
\right]
u(\np)
\\
j_A^\mu(\np',\np)
&=&
\overline{u}(\np')
\left[ G_A \gamma^\mu + G_P\frac{Q^\mu}{2m_N}
\right]\gamma^5
u(\np)\ ,
\end{eqnarray}
where for the isovector nucleon form factors
$F_{1,2}^V= (F_{1,2}^p-F_{1,2}^n)/2$ we use the Galster
parameterization \cite{Gal72} and $u(\np)$ is the free Dirac
spinor of the nucleon. The axial-vector and pseudoscalar form
factors are parameterized as
\begin{eqnarray}
G_A &=& \frac{g_A}{1-Q^2/M_A^2}
\\
G_P &=& \frac{4m_N^2}{m_\pi^2-Q^2}G_A  \label{gp}
\end{eqnarray}
with $g_A=1.26$, $M_A=1032$ MeV.

The SR approximation to this current is then obtained by
inserting the appropriate free spinors $u(\np)$, $u(\np')$, and $\gamma$
matrices, and performing an expansion in powers of $\neta \equiv \np/m_N$ to
first-order.  The procedure was developed in \cite{Ama96a,Ama98b,Ama02b} for
the electromagnetic and transverse neutral, vector and axial-vector 
currents, and is based on the fact that inside the nucleus
$|\neta|$ is a good expansion parameter, the characteristic dimensionless 
nuclear scale being $\eta_F=k_F/m_N\simeq 1/4$.
We exploit QE kinematics, while
further reasonable simplifications are eventually needed in order to arrive at
simple expressions that are easily implementable in traditional non-relativistic
calculations.  For the vector current we use
 the following SR approximation:
\begin{eqnarray}
J_V^0
&=&
\xi_0 +i\xi'_0 (\nkappa\times\neta)\cdot\nsigma
\label{jv0}\\
\nJ^\perp_V
&=&
\xi_1 \neta^\perp+ i\xi'_1 \nsigma\times\nkappa \ ,
\label{jvt}
\end{eqnarray}
where
\begin{eqnarray}
\xi_0 = \frac{\kappa}{\sqrt{\tau}}2G_E^V
&,&
\xi'_0=\frac{2G_M^V-G_E^V}{\sqrt{1+\tau}}
\\
\xi'_1= 2G_M^V  \frac{\sqrt{\tau}}{\kappa}
&,&
\xi_1 = 2G_E^V  \frac{\sqrt{\tau}}{\kappa}
\end{eqnarray}
and use has been made of the dimensionless variables $\nkappa=\nq/2m_N$ and
$\tau=\kappa^2-\lambda^2$.

    From vector current conservation, the longitudinal component is
given by $J^3_V= \frac{\lambda}{\kappa}J^0_V$.  Note that in
Eqs.~(\ref{jv0},\ref{jvt}) the various terms making up the current
are similar to the ones that can be found in traditional
non-relativistic expansions commonly used for the charged current
(see for instance \cite{Krm04}), except for the $\kappa$ and
$\tau$-dependent factors, $\xi_i,\xi'_i$, that provide the
required relativistic behavior.  In $J_V^0$ we include the
first-order ($O(\eta)$) contribution.  This spin-orbit term is
proportional to the operator $(\nkappa\times\neta)\cdot\nsigma$
and is of some importance for high $q$ values.  On the other hand,
the transverse component $\nJ_V^\perp$ is the sum of the usual
magnetization ($\nsigma\times\nkappa$) piece plus a first-order
term, the convection term which is proportional to $\neta^\perp$,
that gives in general a very small \footnote{Note that
$\neta=\np/m_N$ is essentially the velocity $\nv$ of the initial
struck nucleon in units of $c$.} contribution to the cross section
for high $q$ \cite{Ama96a,Jes98}. 

In the case of the axial-vector sector, only the transverse component of
the weak neutral current was expanded in \cite{Ama96a}.
We use the following version of the corresponding SR
current
\begin{equation} \label{jat}
\nJ^\perp_A = \zeta'_1 \nsigma^\perp,
\quad
\zeta'_1 = \sqrt{1+\tau}G_A \ .
\end{equation}
Note that we have neglected the terms of order $\eta$ since, as we
shall show in the next section (and also demonstrated in
\cite{Ama96a} in the context of parity-violating electron
scattering), they are small while adding unnecessary complications
to the shell model calculation.  They can be safely neglected for
our purposes. Remarkably the factor $\sqrt{1+\tau}$ in
Eq.~(\ref{jat}) already accounts for the most part for
relativistic effects in this current (see \cite{Ama96a} for the
full expansion of this current to first-order in $\eta$).

We are left with the 0 and $z$ components of the axial-vector current. Their
SR expressions are presented here for the first time.
Using a notation reminiscent of that used in Appendix~A of \cite{Ama96a},
they are written to first-order in $\eta$ as
\begin{eqnarray}
J_A^0 &=& \zeta'_0 \nkappa\cdot\nsigma +\zeta''_0\neta^\perp\cdot\nsigma
\label{ja0}\\
J_A^z &=& \zeta'_3 \nkappa\cdot\nsigma +\zeta''_3\neta^\perp\cdot\nsigma\ ,
\label{jaz}
\end{eqnarray}
where
\begin{eqnarray}
\zeta'_0 =
\frac{1}{\sqrt{\tau}}\frac{\lambda}{\kappa}G_A'
&,&
\zeta''_0 =
\frac{\kappa}{\sqrt{\tau}}
\left[ G_A -\frac{\lambda^2}{\kappa^2+\kappa\sqrt{\tau(\tau+1)}}G_A'\right]
\\
\zeta'_3 =
\frac{1}{\sqrt{\tau}}G_A'
&,&
\zeta''_3 =
\frac{\lambda}{\sqrt{\tau}}
\left[ G_A -\frac{\kappa}{\kappa+\sqrt{\tau(\tau+1)}}G_A'\right]
\end{eqnarray}
and we have introduced the following combination of axial-vector and pseudoscalar form factors
\begin{equation} \label{gap}
G_A' = G_A-\tau G_P\ .
\end{equation}
For these components of the axial-vector current, Eqs.~(\ref{ja0},\ref{jaz}), we have
performed the expansion to first-order in $\eta$.
The first-order axial-convective term
is proportional to $\nsigma\cdot\neta^\perp$, and, as for
 the spin-orbit and convection terms, only the perpendicular velocity
$\neta^\perp$ appears.
As we shall show
below, for the kinematics of interest in this work the $G_A'$ form factor
which drives the zeroth-order terms turns out to be small due to cancellations in Eq.~(\ref{gap}) between $G_A$ and
$\tau G_P$. In such cases the $O(\eta)$ term is dominant in these current
components.

\subsection{The continuum shell model}

In this work we restrict our attention to the closed-shell
\footnote{By {\em shell} here we mean a sub-shell with quantum numbers
  $(nlj)$.} nuclei $^{12}$C, $^{16}$O and $^{40}$Ca, which we describe in a
Continuum Shell Model (CSM).  
Thus the present model does not include nuclear correlations.  Such
effects are important for low energy, in particular in the axial
responses, and have been estimated for instance in the RPA approach of
Ref. \cite{Nie04}, but are expected to be smaller at the GeV energies
considered in this paper.
The initial state $|i\rangle$ appearing in the hadronic
tensor, Eq.~(\ref{hadronic}), is described as a Slater determinant
representing the uncorrelated nuclear core with all shells occupied.  Since we
are working in the impulse approximation, the final states
are particle-hole excitations coupled to total angular
momentum $J$, namely
$|f\rangle =|(ph^{-1})J\rangle$.  The single hole, $|h\rangle=|\epsilon_hl_hj_h\rangle$, and
particle, $|p\rangle=|\epsilon_pl_pj_p\rangle$, wave functions are obtained by
solving the Schr\"odinger equation with a Woods-Saxon potential
\begin{equation} \label{potencial}
V(r)=-V_0 f(r,R_0,a_0)
+\frac{V_{ls}}{m_\pi^2 r}\frac{df(r,R_0,a_0)}{dr}\nl\cdot\nsigma+V_C(r)\ ,
\end{equation}
where
\begin{equation}
f(r,R,a)=\frac{1}{1+e^{(r-R)/a}}
\end{equation}
and $V_C(r)$ is the Coulomb potential of a charged sphere of
charge $Z-1$ and radius $R_0$ (it is equal to zero for neutrons).  The
parameters of the potential are fitted to the experimental energies
of the valence shells 
and are given in Table~1.
\begin{table}
\begin{center}
\begin{tabular}{ccclllll}
         & $V^p_0$ & $V^p_{LS}$ & $V^n_0$ & $V^n_{LS}$ & $r_0$& $a_0$\\
\hline\hline
$^{12}$C & 62.0 &  3.20  & 60.00 & 3.15   & 1.25 & 0.57 \\
$^{16}$O & 52.5 &  7.00  & 52.50 & 6.54   & 1.27 & 0.53 \\
$^{40}$Ca& 57.5 & 11.11  & 55.00 & 8.50   & 1.20 & 0.53 \\
\hline\hline
\end{tabular}
\caption{\label{tableI}\small
Woods-Saxon potential parameters
for protons (p) and neutrons (n). The units are MeV for $V_i$, and fm
for $a_0$ and $r_0$. The reduced radius parameter $r_0$
is defined by $R_0=r_0 A^{1/3}$.}
\end{center}
\end{table}
In the shell model the energy transfer is computed as the difference between
the (non-relativistic) single-particle energies of particle and hole
$\omega=\epsilon_p-\epsilon_h$.  The relativistic kinematics are taken into
account by the substitution 
\begin{equation} \label{energy}
\epsilon_p\rightarrow
\epsilon_p(1+\epsilon_p/2m_N)
\end{equation}
as the eigenvalue of the Schr\"odinger equation
for the particle; c.f., Eq.~(\ref{kinematics}). 

Since the nuclear states have good angular momentum, it is
convenient for the shell model calculation to perform a multipole expansion of
the components of the current operator in terms of the
usual Coulomb ($C$), longitudinal ($L$),
transverse electric ($E$) and transverse magnetic ($M$)
operators~\cite{Ama96a,Don79,Ama04a}, defined by
\begin{eqnarray}
\hat{C}_{J0}(q) &=&
\int d^3 r j_{J}(qr) Y_{J0}(\hr){J}_0(\nr)
\\
\hat{L}_{J0}(q) &=&
\frac{i}{q}\int d^3 r \nabla\left[j_{J}(qr) Y_{J0}(\hr)\right]
\cdot{\nJ}(\nr)
\label{LJoperator}\\
\hat{E}_{Jm}(q) &=&
\frac{1}{q}\int d^3 r
\nabla\times\left[j_{J}(qr)\nY_{JJm}(\hr)\right]
\cdot{\nJ}(\nr)
\label{EJoperator}\\
\hat{M}_{Jm}(q) &=&
\int d^3 r j_{J}(qr) \nY_{JJm}(\hr)\cdot{\nJ}(\nr) \ ,
\label{MJoperator}
\end{eqnarray}
where $j_J(qr)$ is a spherical Bessel Function and $\nY_{JJm}(\hr)$ is a vector
spherical harmonic.

The nuclear response functions are then written as
\begin{eqnarray}
R_{CC}
&=&
4\pi \sum_{\alpha}
\left(|C^V_\alpha|^2+|C^A_\alpha|^2\right)
\label{rccm}\\
R_{CL}
&=&
2\pi\sum_{\alpha} \left(
C_\alpha^{V*}L_\alpha^V+
C_\alpha^{V}L_\alpha^{V*}+
C_\alpha^{A*}L_\alpha^A+
C_\alpha^{A}L_\alpha^{A*}
\right)
\\
R_{LL} &=&
4\pi\sum_{\alpha}\left(|L^V_\alpha|^2+|L^A_\alpha|^2\right)
\\
R_T
&=&
4\pi\sum_\alpha
\left(|E_\alpha^V|^2+|M_\alpha^V|^2+|E_\alpha^A|^2+|M_\alpha^A|^2
\right)
\\
R_{T'}
&=&
2\pi \sum_\alpha \left(
E_\alpha^{V*}M_\alpha^A+E_\alpha^{V}M_\alpha^{A*}+
E_\alpha^{A*}M_\alpha^V+E_\alpha^{A}M_\alpha^{V*} \right)\ ,
\label{rtpm}
\end{eqnarray}
where in the sums $\epsilon_p=\epsilon_h+\omega$
is modified according to Eq.~(\ref{energy}),
and we use the index $\alpha$ to label the quantum numbers
$\alpha=(h,l_p,j_p,J)$. Moreover,
the $C$, $L$, $E$ and $M$ multipoles of the vector and axial-vector currents
are defined by the reduced matrix elements of the corresponding operators
\begin{eqnarray}
C_\alpha^V+iC_\alpha^A &=& \langle f \| \hat{C}_J(q)\| i\rangle
\\
L_\alpha^V+iL_\alpha^A &=& \langle f \| \hat{L}_J(q)\| i\rangle
\\
E_\alpha^V+iE_\alpha^A &=& \langle f \| \hat{E}_J(q)\| i\rangle
\\
-iM_\alpha^V-M_\alpha^A &=& \langle f \| \hat{M}_J(q)\| i\rangle \ .
\end{eqnarray}
These reduced matrix elements are given in \cite{Ama96a} for the vector
and axial-vector transverse operators, and in \cite{Ama04a} for the
leading-order longitudinal axial-vector component. In the present paper we add the
first-order convective term of the axial-vector current $\neta^\perp\cdot\nsigma$
appearing in Eqs.~(\ref{ja0},\ref{jaz}).  The $C_\alpha^A$ and $L_\alpha^A$
multipoles of this new term are presented in Appendix~A.

The sums over $\alpha$ in Eqs.~(\ref{rccm}--\ref{rtpm}) are infinite
and have to be truncated in the calculation once convergence is reached.
In our approach the number of multipoles is determined by the
maximum of the total angular momentum, $J_{\rm max}$.
We fix this quantity by computing the response functions
setting the potential in the final state to zero and comparing with the
factorized PWIA (see Appendix~B for details). The number $J_{\rm max}$
increases with $q$ and with the number of nucleons, $A$.
In the next section up to 41 multipoles have to be  summed up for 
the case where $q=1.5$ GeV/c.

\section{Results}

\subsection{Test of the SR approach}

\begin{figure}[th]
\begin{center}
\includegraphics[scale=0.95,  bb= 50 560 540 790]{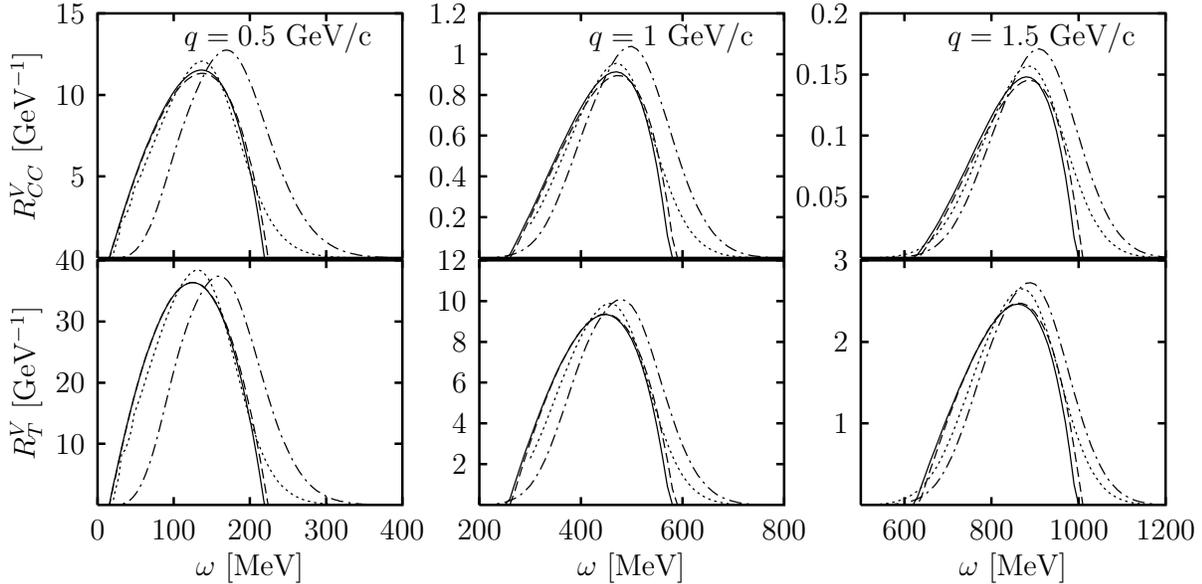}
\caption{Vector response functions of $^{12}$C for three values of the
  momentum transfer. Solid lines: RFG with $k_F=220$ MeV/c.
Dashed lines: SRFG. Dotted: CSM. Dot-dashed: PWIA.
\label{csr1} }
\end{center}
\end{figure}

\begin{figure}[tph]
\begin{center}
\includegraphics[scale=0.95,  bb= 50 350 540 790]{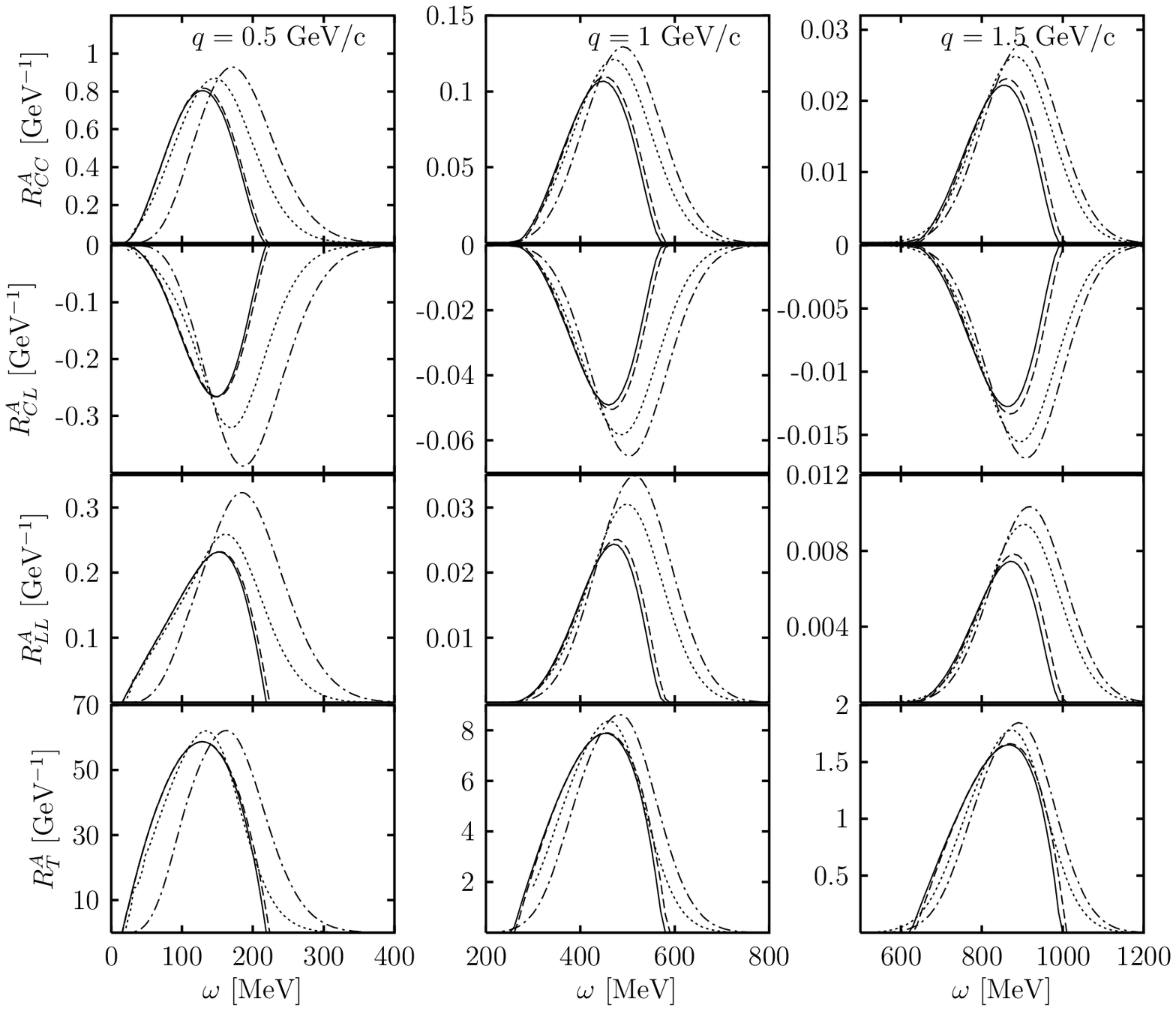}
\caption{As for Fig. 1, but now for the axial-vector responses\label{csr2} }
\end{center}
\end{figure}

\begin{figure}[tph]
\begin{center}
\includegraphics[scale=0.95,  bb= 50 260 540 790]{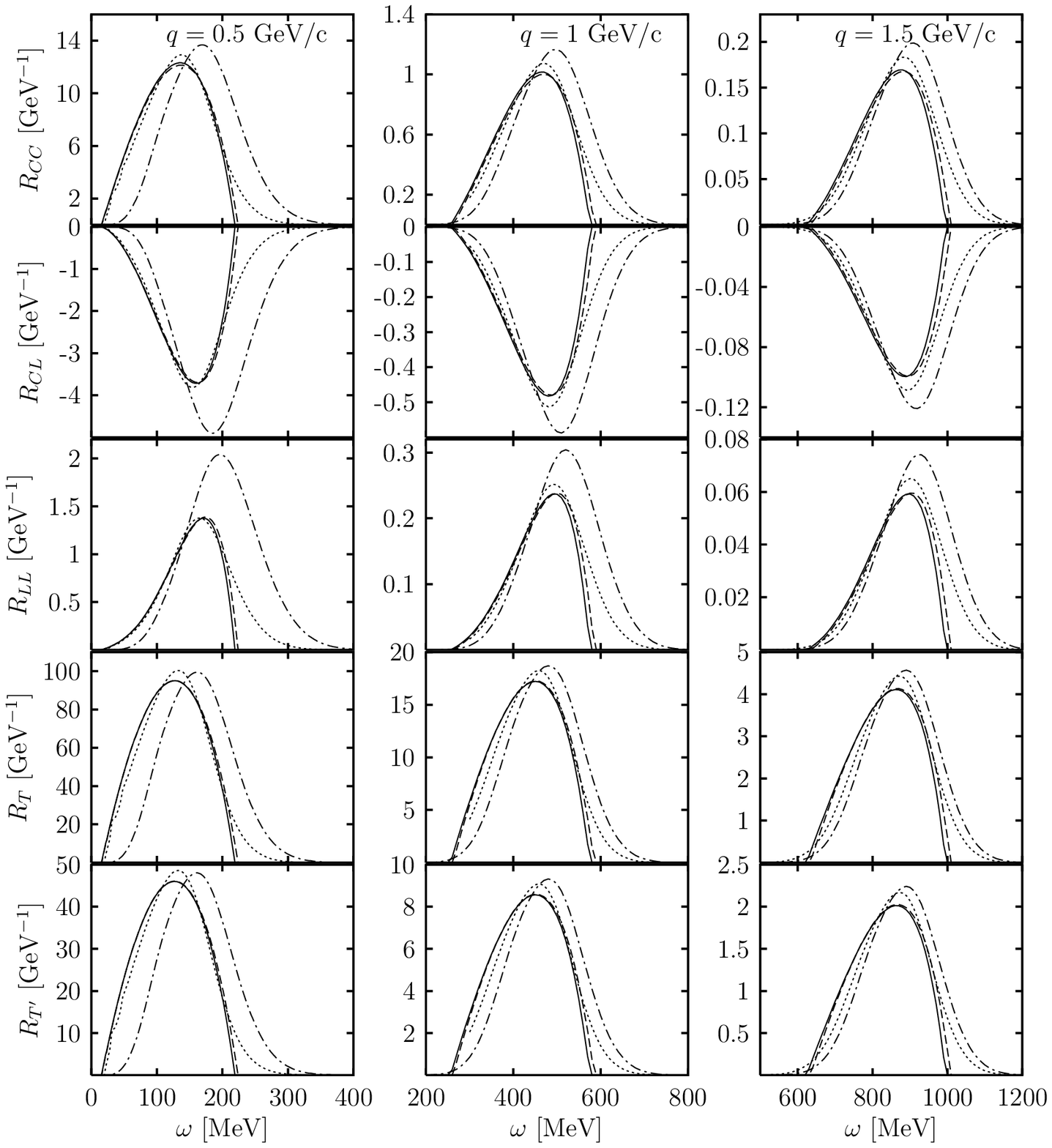}
\caption{As for Fig. 1, but now for the total (vector plus axial-vector) responses
\label{csr3} }
\end{center}
\end{figure}

The quality of the SR expansion of the charged current is illustrated
in Figs.~1--3. There we show the separate vector, axial-vector and
total response functions of $^{12}$C for three values of the momentum
transfer, $q=0.5$, 1, and 1.5 GeV/c.  The Fermi momentum is chosen
to be $k_F=220$ MeV/c.   We show the separate response functions instead
of cross sections because in these functions we can appreciate better
the contribution of the various current components and the behavior of
the SR approximation as a function of the momentum and energy
transfers.  This can be seen in the figures by comparing the solid
lines, representing the RFG result (exact relativistic answer) with
the dashed lines, corresponding to the Semi-Relativistic Fermi Gas
(SRFG). This last model is obtained by implementing the SR expansion of
the current in a non-relativistic Fermi gas, although with
relativistic kinematics. The accord of the two models is almost
perfect for the three values of $q$ considered.  In Fig.~1 we only
show $R_{CC}^V$ and $R_T^V$, since $R_{CL}^V$ and $R_{LL}^V$ are
related to $R_{CC}^V$ by current conservation. In Fig.~2 we show
instead the three response functions, since the axial-vector current
is not conserved.  Finally, in Fig.~3 we show the sums of Figs.~1 and
2 and also the interference response $R_{T'}$.

In Figs.~1--3 we also show with dotted lines the 
CSM responses, computed using the relativizing procedure.  For
comparison we also show the PWIA results obtained by setting 
the potential in the final state to zero (or, equivalently, by integration of
the factorized exclusive responses, as shown in Appendix~B).  The CSM
response functions are quantitatively similar in magnitude to the RFG
and are centered approximately around the same value of $\omega$.  The
major discrepancy between the two models is found for the axial-vector
$R^A_{CC}$, $R^A_{CL}$ and $R^A_{LL}$ response functions, where the
CSM responses are slightly larger in magnitude than the RFG
ones. These responses are in general small compared with the
corresponding vector responses.  This is a consequence of the
suppression of the axial-vector current in the longitudinal channel due to
small value of the form factor $G'_A$, defined in Eq.~(\ref{gap}), for
these kinematics.  In fact, from the definition in Eq.~(\ref{gp}) of the
pseudoscalar form factor one has
\begin{equation} \label{gap2}
G_A'=\left(1-\frac{Q^2}{Q^2-m_\pi^2}\right)G_A \ .
\end{equation}
At the intermediate energies of interest, $|Q^2| \gg m_\pi^2$, and hence
the factor inside the parenthesis in Eq.~(\ref{gap2}) is also small, of
order $O(m_\pi^2/Q^2)$.  In this situation the first-order, axial-vector
convective term of the current, which is proportional to
$\neta^\perp\cdot\nsigma$ (see Eqs.~(\ref{ja0},\ref{jaz})), is dominant over
the zeroth-order contribution, and the corresponding response functions are in
general small compared with the vector ones.  An example is shown in
Fig.~4 where we display the separate contributions of the zeroth- and
first-order terms to the axial-vector response $R_{CC}^A$ of $^{12}$C for
$q=500$ MeV/c in the CSM. For higher values of $q$ the zeroth-order
contribution is much smaller than the others.
\begin{figure}[tph]
\begin{center}
\includegraphics[scale=0.95,  bb= 100 625 500 790]{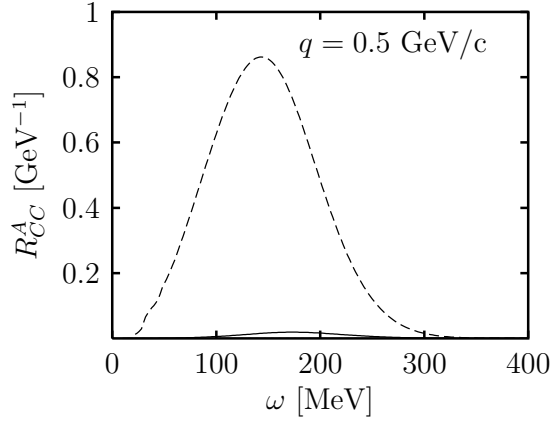}
\caption{Axial-vector response $R_{CC}^A$ of $^{12}$C in the CSM.
Solid: contribution of the zeroth-order term in the SR expansion.
Dashed: first-order contribution. \label{csr4} }
\end{center}
\end{figure}

   From inspection of the responses $R_{CC}$, $R_{T}$ and $R_{T'}$ we
observe that the PWIA results are clearly shifted to the right of the
CSM by roughly the averaged depth of the potential $\sim$35 MeV.  This
shift is present also in the separate vector and axial-vector
responses.  In the case of the $CL$ and $LL$ the shift is larger due
to the energy factors used to compute the $L$ component of the
current.  The origin of the shift in PWIA is related to the different
treatment of the nuclear hamiltonian in the initial and final states
\cite{Ama96b}: while the energy of the initial bound neutron is the
sum of kinetic plus potential, $\epsilon_h=t_h+v_h$, the exiting
proton is a plane wave and has only kinetic energy $\epsilon_p=t_p$.
Accordingly, for fixed $\omega$ there is an imbalance between potential
energies in the initial and final states, yielding $\omega=t_p-t_h-v_h$.  However
in the CSM we use the {\em same} potential for the initial and final
state, which now has $\epsilon_p=t_p+v_p$.  Hence
$\omega=t_p-t_h+v_p-v_h$.  Thus in the CSM the potential energies of
particle and hole partially cancel out and this explains why the
position of the peak is close to the RFG, where only kinetic energies
enter. Of course the cancellation is not perfect, and a slight shift
to the right of the RFG is also observed in the CSM.
This small shift of the CSM is the behavior expected from previous 
theoretical studies \cite{Ros78}.

\begin{figure}[t]
\begin{center}
\includegraphics[scale=0.9,  bb= 50 470 540 805]{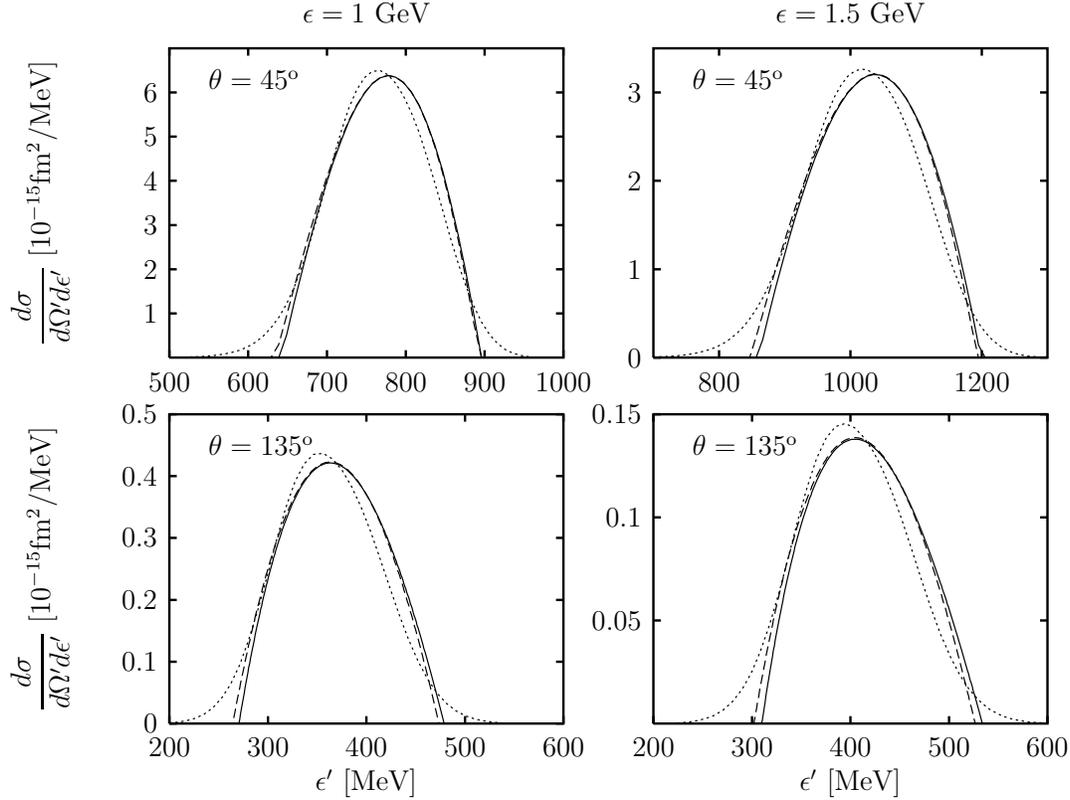}
\caption{Differential cross section of the reaction
  $^{12}{\rm C}(\nu_\mu,\mu^-)$ for incident neutrino energies $\epsilon=1$
  and 1.5 GeV and for two scattering angles.  Solid lines: RFG with $k_F=215$
  MeV/c.  Dashed lines: SRFG.  Dotted: CSM.
\label{csr5} }
\end{center}
\end{figure}

Another test of the SR approximation is illustrated by the results
shown in Fig.~5 for the differential cross section $^{12}{\rm
C}(\nu_\mu,\mu^-)$. Therein we show examples for two incident
neutrino energies, $\epsilon=1$ and 1.5 GeV, and for two
scattering angles $\theta=45^{\rm o},135^{\rm o}$, as a function
of the exiting muon energy $\epsilon'$; so in this case we are
testing different ranges of $q$ and $\omega$, which can be high or
low depending on the kinematics.  We see again that the RFG and
SRFG predictions are almost equal in all the cases, while the CSM
also gives similar results, with the exception of the
characteristic tails and slight shift, that now is to the left of
the RFG since $\omega$ decreases with $\epsilon'$.

Summarizing this subsection, using the Fermi gas as ``testing arena''
for the approximated CC current, we have found that the accord between
the results obtained using the SR expansion and the exact relativistic
result is almost perfect for the intermediate and high values of $q$ and
$\omega$ considered.  
The CSM relativized using our procedure gives
rise to cross sections which are within the allowed kinematical region
and of a magnitude similar to those of the RFG.
This behavior of the CSM is similar to what was found 
for $(e,e')$ reactions in \cite{Ama96a}. 
We should underscore the fact that the CSM is used here just as illustration of  
how one can use our relativizing procedure in a more complex model than the
Fermi gas, and of the  results that are to be expected from it. In particular,
especially when discussing lower energies where non-relativistic approaches are valid, 
more elaborated models exist that describe the $(e,e')$ experimental data 
rather well. For example, it is well known from other work that final-state 
interactions can significantly modify the bare CSM or RPA responses, mainly through 
medium renormalization of the particle-hole excitations. 
This mechanism can be approximately taken into account by using an effective
nucleon mass that produces a shift and by a folding of the bare responses,
producing an asymmetric broadening which improves the agreement with 
the experimental data \cite{Smi88,Ama93,Ama94}.

\subsection{Scaling}

In this subsection we work within the CSM. We first focus on the
inclusive electron scattering reaction, and investigate the scaling
properties of the electromagnetic responses for various choices of kinematics
for $^{12}$C, $^{16}$O, and $^{40}$Ca.  The
$(e,e')$ cross section is given by
\begin{equation}
\frac{d\sigma}{d\Omega' d\epsilon'} = \sigma_{Mott}
\left( v_L R_L + v_T R_T \right)\ .
\end{equation}
The same SR expansion for the vector current,
Eqs.~(\ref{jv0},\ref{jvt}),  is employed
here for the electromagnetic sector.

We compute the CSM scaling functions as
\begin{eqnarray}
f_L &=& \frac{R_L}{G_L}
\\
f_T &=& \frac{R_T}{G_T}
\end{eqnarray}
with
\begin{equation}
G_K = \Lambda_0 \left( ZU_K^{p}+N U_K^n\right) \quad\quad K=L,T,
\end{equation}
where $\Lambda_0$ is given in Eq.~(\ref{lambda0}) and
the electromagnetic single-nucleon functions for protons,
$U_K^p$, and neutrons, $U_K^n$, are defined similarly to the ones given
in Appendix~B for the vector current, $U_{CC}^V$ and $U_T^V$.

The scaling behavior can be studied by plotting these functions
against the scaling variable~\cite{Alb88}
\begin{equation} \label{psi}
\psi=\frac{1}{\sqrt{\xi_F}}
\frac{\lambda-\tau}{\sqrt{(1+\lambda)\tau+\kappa\sqrt{\tau(1+\tau)}}}
\end{equation}
for different kinematics and for different nuclei.
Here $\xi_F=\sqrt{1+k_F^2/m_N^2}-1$.

\begin{figure}[tph]
\begin{center}
\includegraphics[scale=0.9,  bb= 100 360 500 790]{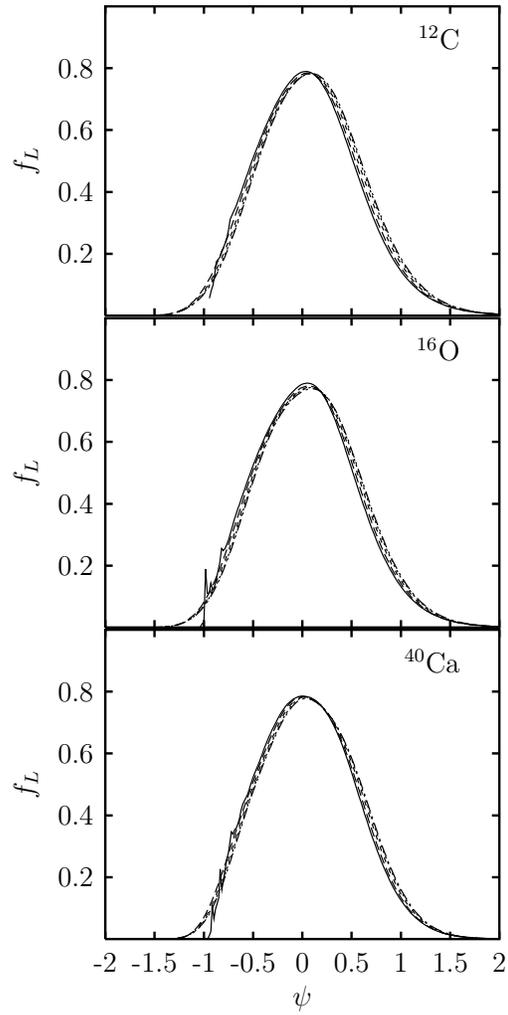}
\caption{Scaling of the first kind in the CSM with scaling functions
obtained from the longitudinal electromagnetic responses in the
CSM. In each panel we include $q=0.5$, 0.7, 1, 1.3 and 1.5 GeV/c.
\label{sca1}
}
\end{center}
\end{figure}

A study of the behavior of $f_L$, computed for various values of the
momentum transfer, is summarized in Fig.~\ref{sca1}.  We plot together
$f_L$ for $q=0.5$, 0.7, 1, 1.3, and 1.5 GeV/c. All of the curves
approximately collapse into one.  Small violations of the scaling are
seen at low $\psi$ coming from the low-energy potential resonances for
$q=0.5$ GeV/c, which disappear for higher $q$-values. Thus scaling of
the first kind is approximately achieved in the CSM.
Note that scaling also holds for $|\psi|>1$, the region
where the RFG responses are zero.

\begin{figure}[tp]
\begin{center}
\includegraphics[scale=0.9,  bb= 150 320 440 780]{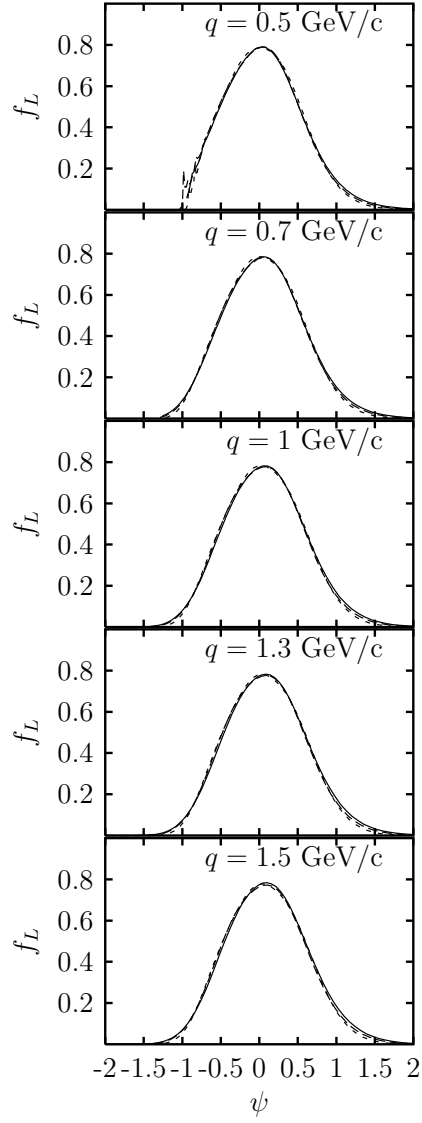}
\caption{As for Fig.~6, but now including in each panel the nuclei
$^{12}$C, $^{16}$O and $^{40}$Ca. \label{sca2}}
\end{center}
\end{figure}

\begin{figure}[tp]
\begin{center}
\includegraphics[scale=0.9,  bb= 150 320 440 805]{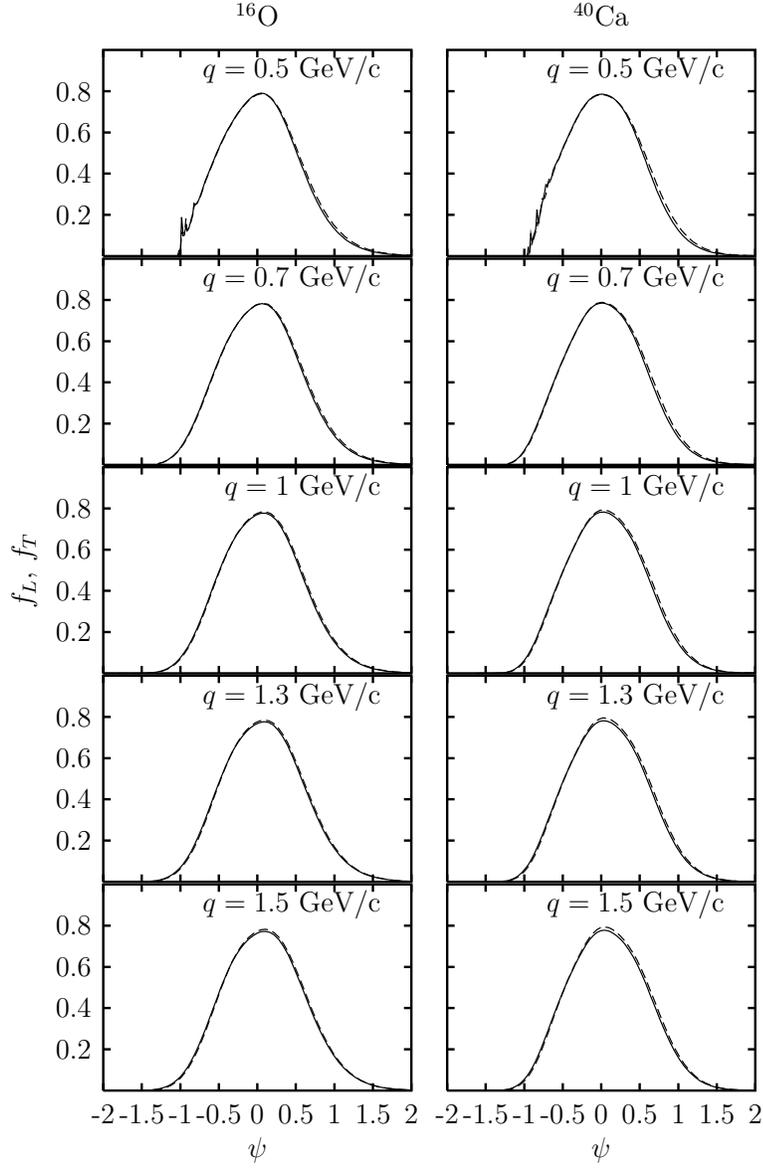}
\caption{Scaling function $f_L$ obtained from
the longitudinal electromagnetic response (solid lines),
compared with the scaling function $f_T$ obtained from the transverse response
(dashed lines), for several values of $q$.
\label{sca3}}
\end{center}
\end{figure}

Scaling of the second kind, {\it i.e.,} independence of the nuclear species
for fixed $q$, is illustrated in Fig.~\ref{sca2}. We plot together
$f_L$ for the three nuclei studied. 
The fitted Fermi momenta are 220, 215 and 240 MeV/c for 
$^{12}$C, $^{16}$O and $^{40}$Ca, respectively. 
The collapse of the three curves
into one is clearly seen, with small deviations at the region of
the maximum.  Again the exception is found for $q=0.5$ GeV/c in the
resonance region.  
Since both kinds of scaling are found, we conclude
that superscaling occurs within our model.

Finally, in Fig.~8 we show what has been called scaling of the zeroth
kind \cite{Don99b}, {\it i.e.,} the longitudinal and transverse
scaling functions, $f_L$ and $f_T$, also collapse into one universal
function $f$.  Experimentally, deviations from scaling in the region
of the QE peak mainly occur in
the transverse response, and are related to contributions beyond the
impulse approximation, in particular to meson-exchange currents
\cite{Don99a,Don99b}. Since these contributions are not included in
our CSM, the $L$ and $T$ responses scale in the same way, as shown in
Fig.~\ref{sca3}.

In the present work we do not compare with the experimental $(e,e')$ scaling
function because, as stated above, the CSM is still lacking some
ingredients (for instance, medium modifications to the p-h propagator,
inclusion of energy-dependent potentials, {\it etc.}), without which one should
not expect to obtain excellent agreement with the experimental
data.  The focus of the present study has been limited to
showing that superscaling occurs at the level of relativized CSM.

Let us now turn to neutrino reactions. In \cite{Ama05b} a
semi-empirical model, based on the superscaling property of the $L$
responses of $^{12}$C and $^{16}$O for high momentum transfer, was
proposed to predict the neutrino inclusive cross sections up to the
$\Delta$ peak.  The latter were calculated starting from the RFG
expression for the cross sections and substituting the ``theoretical''
RFG superscaling functions with phenomenological ones (one for the QE
and a different one for the $\Delta$ peak), derived from fits of
electron scattering data.  Therefore this reconstruction, besides
assuming the validity of superscaling in electron scattering, relies
on the hypothesis that it also holds for the neutrino inclusive cross
section for the high energies involved, so that electron scattering
results can safely be used as input in neutrino scattering
calculations.  This assumption obviously cannot be tested using the
RFG model, since, as stated in the introduction, it is true by
construction.

In the present work we have at hand a model, the CSM, for both the QE
$(e,e')$ and $(\nu_\mu,\mu^-)$ reactions that has been relativized 
and so should be able to handle modeling at high energies.  Moreover, superscaling
of the electromagnetic responses is well satisfied by the CSM.  Hence
we can adopt the same approach as in \cite{Ama05b} treating the CSM
electromagnetic scaling function as a pseudo-phenomenological one and
using it to compute neutrino cross sections. Upon comparing cross
sections obtained this way with those obtained directly using the CSM
we can then gain some insight into the degree to which the shell model
incorporates effects which are scale-breaking. While the level of
scaling-violation is expected to be quite small, since we have already
seen excellent superscaling in the figures discussed above, the
specific differences in the roles played by the various current
operators involved in electromagnetic and CC weak processes might lead
to different sensitivities to scale-breaking effects.  Accordingly, it
is useful to compare the cross sections obtained within the scaling
approach with those computed directly using the model. In context it
should be noted that this does not imply in general that the scaling
approach has been shown to be robust, since here the comparisons are
being made entirely within limited models which clearly lack elements
that may play some role in the responses and may be scale-breaking ---
we know of at least one such ingredient, namely, meson-exchange
currents which have been shown to provide scale-breaking corrections that
enter typically at the 10\% level in the overall cross
sections. Nevertheless, the study presented below indicates that the
scale-breaking effects incorporated in the present models are in fact 
quite small.

\begin{figure}[t]
\begin{center}
\includegraphics[scale=0.9,  bb= 50 470 540 805]{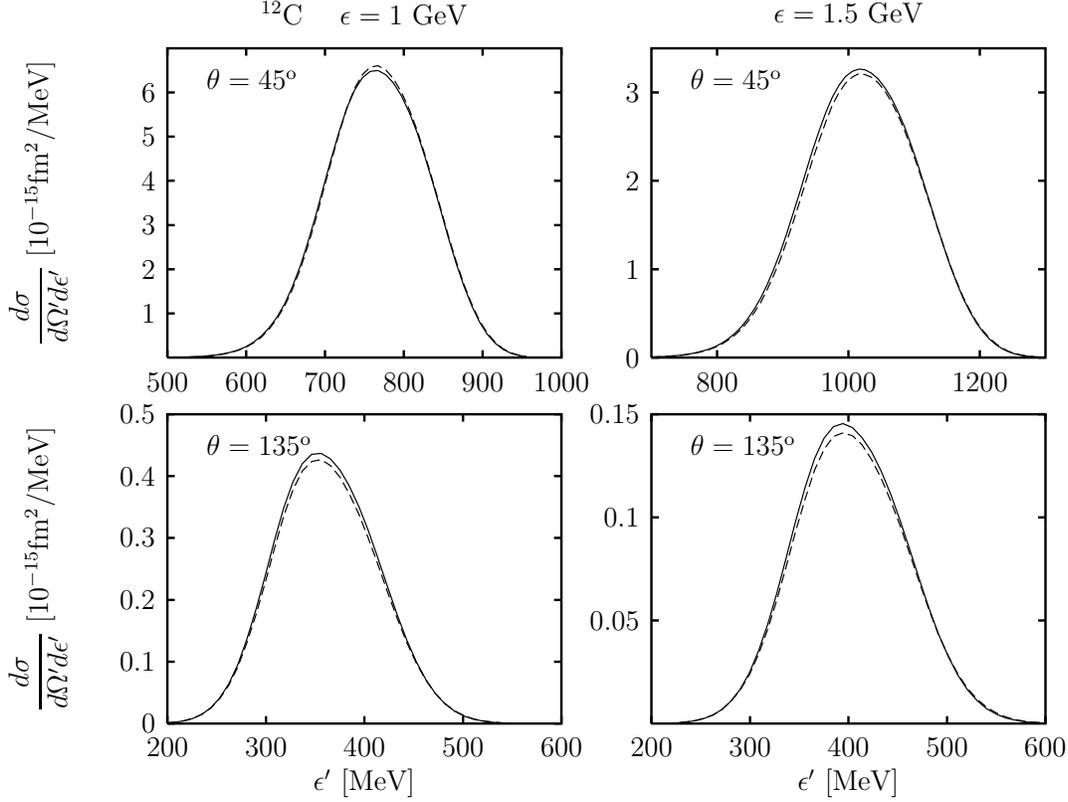}
\caption{Differential cross section of the reaction
  $^{12}{\rm C}(\nu_\mu,\mu^-)$ for neutrino incident energies $\epsilon=1$
  and 1.5 GeV and for two scattering angles.  Solid lines: CSM.
  Dashed lines: reconstructed
from the electromagnetic scaling function $f_L(\psi)$
computed at the same kinematics.
\label{sca4} }
\end{center}
\end{figure}

\begin{figure}[t]
\begin{center}
\includegraphics[scale=0.9,  bb= 50 470 540 805]{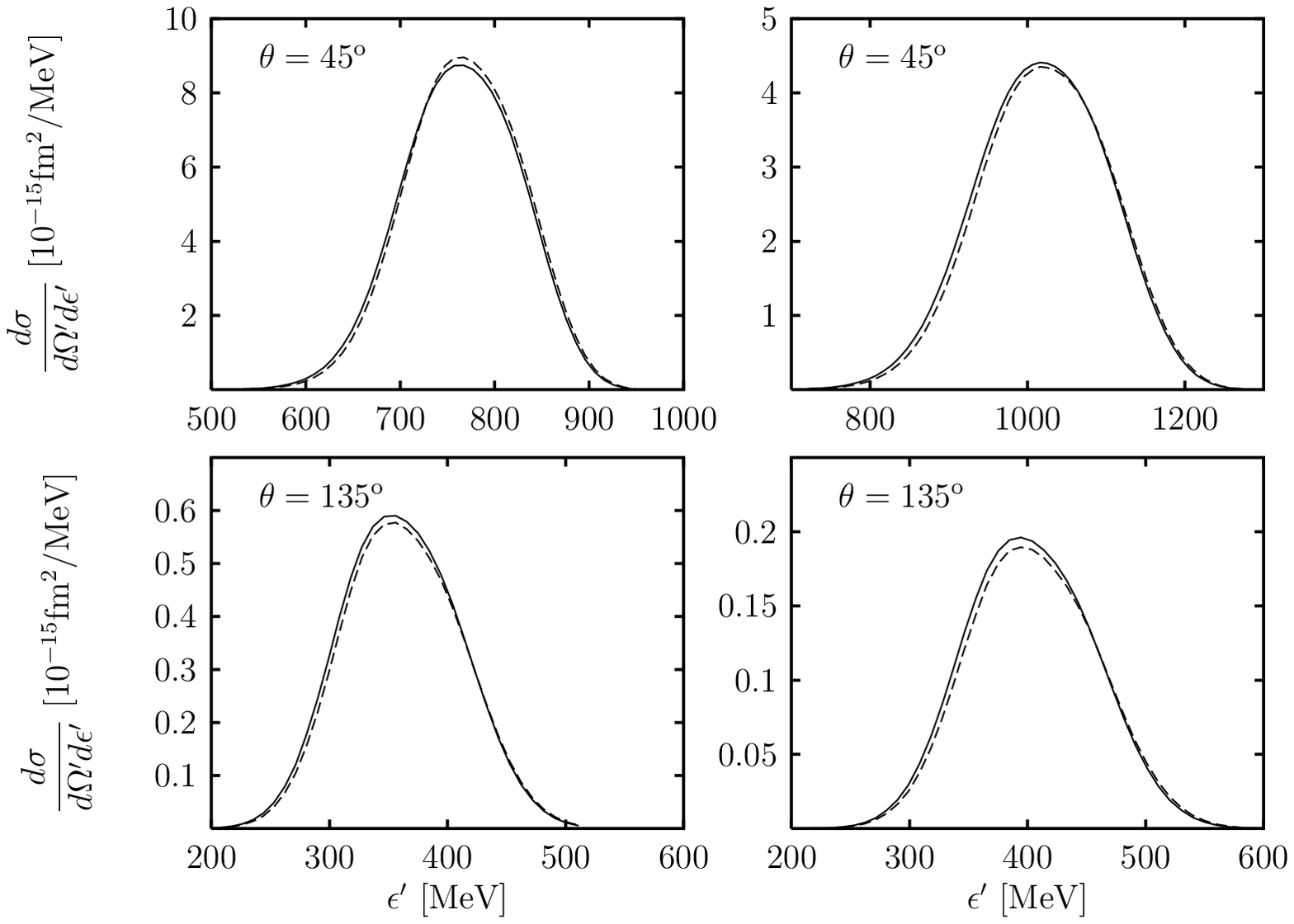}
\caption{As for Fig. 9, but now for $^{16}$O.
\label{sca5} }
\end{center}
\end{figure}

We show typical results of this study in Figs.~\ref{sca4} and
\ref{sca5} for $^{12}$C and $^{16}$O, respectively.  The cross section
for $(\nu_\mu,\mu^-)$ is shown for two incident energies and for two
scattering angles.  The dashed lines have been computed from the
scaling function $f_L$ obtained from the analysis of $(e,e')$ CSM predictions.
To be precise, we use Eq.~(\ref{rfg}) to
compute the neutrino responses, substituting $f^{RFG}$ by $f_L$.  The
solid lines correspond to the exact CSM result.  The differences found between
the two approaches are quite small, at most $\sim 3\%$.  These small
differences are produced by the slightly different scaling behavior of
the axial-vector and vector responses, that, however, has little effect on the
total cross section. Due to the simplicity of the CSM, these results cannot 
be taken as a definite proof of superscaling, but certainly represent an important
step forward in establishing the validity of the approach presented in
\cite{Ama05b}.

\section{Conclusions}

One of the goals of the present study has been to explore the
degree to which scaling and superscaling behaviors are reached
for relatively high-energy semi-leptonic inclusive reactions with
nuclei at excitation energies in the vicinity of the quasielastic peak.
The focus has been placed on comparisons between the Fermi gas model
(both the fully relativistic Fermi gas and a semi-relativized version
of it) and a relativized continuum shell model. 

Since we are interested
in energies of several GeV, relativity is clearly a required ingredient.
In the present work we have presented a clear and direct procedure that allows one to
incorporate some classes of relativistic effects and thereby to
relativize otherwise non-relativistic models of the reaction, such as the
shell model employed here. We use the two versions of the Fermi gas model
to motivate the procedures followed. Two steps are involved: first, one
relativizes the kinematics in the reaction, and second, expansions are made
for the single-nucleon currents in the problem. In the present work we have
extended our previous treatments to include the full charge-changing weak
interaction current. Importantly, because the expansions are made to first-order 
in $p/m_N$, but not at all in $q/m_N$ or $\omega/m_N$, they differ from the 
traditional FW expansions and for
QE scattering should be more robust at high energies where
expansions in $q/m_N$ and $\omega/m_N$ clearly fail.

Using the two versions of the Fermi gas model and the relativized continuum
shell model, together with some results from the plane-wave impulse approximation, 
to represent both electromagnetic, $(e,e')$, and CC weak, $(\nu_\mu,\mu^-)$, 
processes, we have proceeded to quantify the level of scaling violation. 
This allows us to evaluate the corresponding uncertainty in the
predictions one makes for the neutrino cross section using the
scaling approach with input from analyses of electron scattering data.  
In the present paper we have shown that for
intermediate to high energies the description of the final state
through a real mean field seems not to change appreciably the
scaling and superscaling properties of the electromagnetic
response functions at the quasielastic peak, and that the scaling
function extracted from these is basically the same as for the
$(\nu_\mu,\mu^-)$ reaction.  

Although the present analysis performed within the relativized shell model does not
constitute a proof of the scaling hypothesis in the general case, scaling
studies such as those presented here allow us to gain additional insight 
and, in particular, to rule
out (or not) some reaction mechanisms as possible causes of scaling violations.

In summary, the relativizing procedure followed in the present work embodies
some of the ingredients that are almost certainly required for a sucessful
description of such inclusive processes in the quaselastic regime --- models
which do not take at least these ingredients into account are very likely to
fail in the several GeV energy region of interest here. This does not mean,
however, that additional dynamical features are not needed before a full
understanding of these high-energy responses will be attained. In particular,
other studies being pursued in parallel by us hold some promise for reaching a good
understanding of the phenomenologically-derived scaling function used recently
to make predictions for CC neutrino reactions in the GeV region. Some of these
results will be presented in the near future.

\section*{Acknowledgments}

This work was partially supported by funds provided by DGI (Spain) and
FEDER funds, under Contracts Nos. BFM2002-03218, BFM2002-03315 and
FPA2002-04181-C04-04 and by the Junta de Andaluc\'{\i}a,
and by the INFN-CICYT collaboration agreement (project
``Study of relativistic dynamics in electron and neutrino scattering'').
It was also supported in part (TWD) by the U.S. Department of Energy
under cooperative research agreement No. DE-FC02-94ER40818.

\appendix

\section{Multipoles of the convective axial-vector operator}

Here we compute the Coulomb multipoles of the convective axial-vector charge
operator,
which is the first-order term in the expansion of $j_A^0$ (Eq.~(\ref{ja0})).
The corresponding coordinate-space operator is given by
\begin{equation}
\rho_C(\nq) \equiv e^{i\nq\cdot\nr} \zeta''_0\,\neta_\perp\cdot\nsigma \ ,
\end{equation}
where $\nr$ is the coordinate of the nucleon over which the operator acts.
To obtain the multipole operators we proceed by writing the above operator in the
form:
\begin{eqnarray}
\rho_C(\nq)
&=&
 e^{i\nq\cdot\nr} \zeta''_0\,
\left(\neta-\frac{\neta\cdot\nq}{q^2}\nq\right)\cdot\nsigma
\\
&=&
\zeta''_0\,
\nsigma\cdot \left(\neta+\frac{\neta\cdot\nabla}{q^2}\nabla\right)
 e^{i\nq\cdot\nr} \ .
\end{eqnarray}
Performing in this equation the multipole expansion of the
plane wave for $\nq$ along the $z$-axis
\begin{equation}
e^{i\nq\cdot\nr}=\sqrt{4\pi}\sum_J i^J [J]j_J(qr)Y_{J0}(\hr),
\end{equation}
where $[J]=\sqrt{2J+1}$, we obtain
\begin{equation}
\rho_C(\nq) = \sqrt{4\pi}\sum_{J} i^J [J] \hat{C}_{J0}(q)\ ,
\end{equation}
where the Coulomb multipole operators are
\begin{equation}
\hat{C}_{J0}(q) =
\zeta''_0\,
\nsigma\cdot \left(\neta+\frac{\neta\cdot\nabla}{q^2}\nabla\right)
j_J(qr)Y_{J0}(\hr)\ .
\end{equation}
Later on we will also make the substitution $\np \rightarrow -i\nabla$,
although in this equation $\nabla$ operates only on
$j_J(qr)Y_{J0}(\hr)$.

The procedure now is to use Racah algebra and the  general properties of
spherical harmonics and spherical Bessel functions to write the above operator
in explicit spherical-tensor form. The following expressions can be obtained:
\begin{eqnarray}
\hat{C}_{J0}
&=&
\zeta''_0 \left[\hat{C}_{J0}^{(1)} +
\hat{C}_{J0}^{(2)}\right]
\\
\hat{C}_{J0}^{(1)}
&=&
\frac{i}{m}\frac{1}{[J]}
\sum_L [L] U_{JLJ}
\label{cj1}
\\
\hat{C}_{J0}^{(2)}
&=&
-\frac{i}{m}
\sum_{s=\pm 1}\sum_{s'=\pm 1}
\frac{(J+\delta_{s1})^{1/2}(J'+\delta_{s'1})^{1/2}}{[J][J']}
U_{J''J'J}\ ,
\label{cj2}
\end{eqnarray}
where $J'=J+s$, $J''=J'+s'$ and $s,s'=\pm1$ (coming from the
derivative  of the Bessel function, which is a linear combination
of the Bessel functions for $J\pm1$), and we have defined the
auxiliary coupled operator
\begin{equation}
U_{J''J'J} \equiv j_{J''}(qr)
\left[\sigma\otimes\left[Y_{J''}(\hr)\otimes\nabla\right]_{J'}\right]_{J0}\ .
\end{equation}

Next we proceed  to compute the reduced matrix elements of this $U$-operator
between shell-model particle and hole states
$|p\rangle=|\frac12l_pj_p\rangle$, and $|h\rangle=|\frac12l_hj_h\rangle$.
Using standard Racah algebra
\cite{Edm74,Rot59} we obtain
\begin{eqnarray}
\langle p \| U_{J''J'J} \| h\rangle
&=&
(-)^{l_h+J'}\sqrt{\frac{3}{2\pi}}
[l_p][j_p][j_h][J][J'][J'']
\ninej{\frac12}{l_p}{j_p}{\frac12}{l_h}{j_h}{1}{J'}{J}
\nonumber\\
&&\mbox{}\times
 \sum_{s_h=\pm 1} [L_h]s_h (l_h+\delta_{s_h1})^{1/2}
 \sixj{J''}{1}{J'}{l_h}{l_p}{L_h}
\threej{l_p}{J''}{L_h}{0}{0}{0}
\nonumber\\
&&\mbox{}\times
\int_0^\infty dr\, r^2 R_p(r)j_{J''}(qr)
\left(\frac{d}{dr}-s_h\frac{l_h+\delta_{s_h,-1}}{r}\right) R_h(r)\ ,
\end{eqnarray}
where $R_p(r)$ and $R_h(r)$ are the radial wave functions,
$L_h=l_h+s_h$, and $s_h=\pm 1$.
We can now use this result in Eqs.~(\ref{cj1},\ref{cj2}).
It is convenient to use the following
identities involving products of six-j and three-j coefficients,
for $J''=J'+s'$ and $L_h=l_h+s_h$, with $s',s_h=\pm1$:
\begin{eqnarray}
\sixj{J''}{1}{J'}{l_h}{l_p}{L_h}
\threej{l_p}{J''}{L_h}{0}{0}{0}
&=&
\frac{P^+_{l_p+l_h+J'}}{[L_h][l_h][J'][J'']}
\nonumber\\
&&
\left\{
   \left[
     \left(l_h+\delta_{s_h,-1}\right)
     \left(J'+\delta_{s',-1}\right)
   \right]^{1/2}
   \threej{l_h}{J'}{l_p}{1}{-1}{0}
\right.
\nonumber\\
&&\mbox{}
\kern -1cm
-\left.
   s_h s'
   \left[
     \left(l_h+\delta_{s_h,1}\right)
     \left(J'+\delta_{s',1}\right)
   \right]^{1/2}
   \threej{l_h}{J'}{l_p}{0}{0}{0}
\right\}
\\
\sixj{J}{1}{J}{l_h}{l_p}{L_h}
\threej{l_p}{J}{L_h}{0}{0}{0}
&=&
\frac{P^+_{l_p+l_h+J+1}}{[L_h][l_h][J]}
     \left(l_h+\delta_{s_h,-1}\right)
   \threej{l_h}{J}{l_p}{1}{-1}{0}
\end{eqnarray}
where $P^+_n$ is the parity function equal to one if $n$ is even
and zero if $n$ is odd.
We also use  the following product of a nine-j and a three-j
\begin{eqnarray}
\ninej{\frac12}{l_p}{j_p}{\frac12}{l_h}{j_h}{1}{J'}{J}
\threej{l_h}{J'}{l_p}{0}{0}{0}
&=&
\frac{(-)^{j_p+l_p+\frac12}}{\sqrt{6}}
\frac{P^+_{l_p+l_h+J+1}}{[l_p][l_h][J][J']}
\nonumber\\
&& \mbox{}\times
\frac{\chi_p+\chi_h+sJ+\delta_{s1}}{\sqrt{J+\delta_{s1}}}
\threej{j_p}{j_h}{J}{\frac12}{-\frac12}{0}\ ,
\end{eqnarray}
where $\chi_p=(-1)^{l_p+j_p+\frac12}(j_p+\frac12)$.
After some work, we finally arrive at the matrix elements written in the form
\begin{eqnarray}
\langle p \|  \hat{C}_J^{(1)}\| h\rangle
&=&
\frac{i}{m}A_J(ph)\int_0^\infty dr\, r R_p^*(r)j_J(qr)R_h(r)
\nonumber\\
&&\mbox{}+\frac{i}{m}
B_J(ph)\int_0^\infty dr\, r^2 R_p^*(r)j_J(qr)\frac{dR_h(r)}{dr}
\\
\langle p \|  \hat{C}_J^{(2)}\| h\rangle
&=&
\frac{i}{m}\sum_{s=\pm1}
 A_{JJ'}(ph)\frac{1}{q}\int_0^\infty dr\, R_p^*(r)j_{J'}(qr)R_h(r)
\nonumber\\
&&\mbox{}+\frac{i}{m}\sum_{s=\pm1}
B_{JJ'}(ph)\int_0^\infty dr\, r^2 R_p^*(r)j'_{J'}(qr)\frac{dR_h(r)}{dr}\ ,
\end{eqnarray}
where $J'=J+s$, $s=\pm1$,
 and $j'_{J'}(z)$ is the derivative of the Bessel
function. We have defined the following coefficients
\begin{eqnarray}
A_J(ph)&=&
P^+_{l_p+l_h+J+1}
\left[  [J]\sqrt{l_h(l_h+1)} b_{JJ}
      -\sum_{s=\pm1}\frac{[J']}{[J]}\sqrt{J'+\delta_{s1}}b_{JJ'}
\right]
a_{J}
\\
B_J(ph)&=&
P^+_{l_p+l_h+J+1}
\frac{(-)^{j_p+\frac12}}{\sqrt{4\pi}}[j_p][j_h][J]
\threej{j_p}{j_h}{J}{\frac12}{-\frac12}{0}
a_{J}
\\
A_{JJ'}(ph)&=&
P^+_{l_p+l_h+J+1}
\sqrt{J+\delta_{s1}}
\sqrt{l_h(l_h+1)}\sqrt{J'(J'+1)}
a_{J'}b_{JJ'}
\\
B_{JJ'}(ph)&=& P^+_{l_p+l_h+J+1} \frac{(-)^{j_p-1/2}}{\sqrt{4\pi}}
\frac{[j_p][j_h]}{[J]} \left(\chi_p+\chi_h+sJ+\delta_{s1}\right)
\threej{j_p}{j_h}{J}{\frac12}{-\frac12}{0},
\end{eqnarray}
where the factors $a_J$ and $b_{JJ'}$ are defined as follows:
\begin{eqnarray}
a_{J}&=& (-)^{l_p}\sqrt{\frac{3}{2\pi}}[l_p][l_h][j_p][j_h][J]
\\
b_{JJ'}&=& \ninej{\frac12}{l_p}{j_p}{\frac12}{l_h}{j_h}{1}{J'}{J}
                \threej{l_h}{J'}{l_p}{1}{-1}{0}.
\end{eqnarray}

\section{PWIA}

We follow the approach of \cite{Ama96b}.
In PWIA  the inclusive response functions can be written as an integral
over the missing momentum $\np=\np'-\nq$, with $p'=\sqrt{2m_N\epsilon_p}$,
\begin{equation}\label{rpwia}
\left[R_K^{PWIA}(q,\omega)\right]_h
= \frac{m_N}{q}\int_{|p'-q|}^{p+q}dp \, p \int_0^{2\pi} d\phi\,
w_K(\np',\np)M_h(\np)
\end{equation}
of the scalar momentum distribution for each occupied shell
\begin{equation}
M_h(\np) = \frac{2j_h+1}{4\pi}|\tilde{R}_h(p)|^2\ ,
\end{equation}
where $\tilde{R}_h(p)$ is the radial wave function in momentum space.
The single-nucleon exclusive
responses $w_K(\np',\np)$, for $K=CC,CL,LL,T,T'$
 are readily computed using the vector,
Eqs.~(\ref{jv0},\ref{jvt}), and axial-vector, Eqs.~(\ref{jat},\ref{ja0},\ref{jaz}),
current components
\begin{eqnarray}
w_{CC} &=& w_{CC}^{V}+w_{CC}^A \\
w_{CC}^V &=& \xi_0^2+\xi'_0{}^2\kappa^2\eta_\perp^2\\
w_{CC}^A &=& \zeta'_0{}^2\kappa^2+\zeta''_0{}^2\eta_\perp^2 \\
w_{CL} &=& w_{CL}^{V}+w_{CL}^A \\
w_{CL}^V &=& -\frac{\lambda}{\kappa}w_{CC}^V \\
w_{CL}^A &=& -\zeta'_0\zeta'_3\kappa^2-\zeta''_0\zeta''_3\eta_\perp^2 \\
w_{LL} &=& w_{LL}^{V}+w_{LL}^A \\
w_{LL}^V &=& \left(\frac{\lambda}{\kappa}\right)^2 w_{CC}^V \\
w_{LL}^A &=& \zeta'_3{}^2\kappa^2+\zeta''_3{}^2\eta_\perp^2 \\
w_{T} &=& w_{T}^{V}+w_{T}^A \\
w_{T}^V &=& 2\xi'_1{}^2\kappa^2+\xi_1^2\eta_\perp^2 \\
w_{T}^A &=& 2\zeta'_1{}^2 \\
w_{T'} &=& 2\xi'_1\zeta'_1\kappa\ .
\end{eqnarray}
Since these functions do not depend on the azimuthal angle
of $\np$, namely $\phi$, the response functions in Eq.~(\ref{rpwia}) are reduced to an
integral over the missing momentum $p$, which is performed numerically.

\section{The relativistic Fermi gas}

Here we summarize the expressions for $(\nu_\mu,\mu^-)$ reactions in the RFG
model.  We follow \cite{Ama05a}, where the expressions were written to
leading order. Here we write the full results for the non-Pauli blocked regime
of interest in this work.
 The nuclear response functions
are written as
\begin{equation} \label{rfg}
R_K = N \Lambda_0 U_K  f_{RFG}(\psi),  \quad K=CC,CL,LL,T,T',
\end{equation}
where $N$ is the neutron number,
\begin{equation}\label{lambda0}
\Lambda_0 = \frac{\xi_F}{m_N \eta_F^3 \kappa}\ .
\end{equation}
Here $\eta_F=k_F/m_N$, $\xi_F=\sqrt{1+\eta_F^2}-1$.
In Eq.~(\ref{rfg}) $f_{RFG}(\psi)$ is the scaling function of the RFG
\begin{equation}
f_{RFG}(\psi)=\frac34 (1-\psi^2)\theta(1-\psi^2)
\end{equation}
and $\psi$ is the scaling variable
given in Eq.~(\ref{psi}).
Finally, we give the single-nucleon responses $U_K$.
For $K=CC$ we have
\begin{eqnarray}
U_{CC} &=& U_{CC}^V+
\left(U_{CC}^A\right)_{\rm c.}
+\left(U_{CC}^A\right)_{\rm n.c.}
\\
U_{CC}^V &=&
\frac{\kappa^2}{\tau}
\left[ (2G_E^V)^2+\frac{(2G_E^V)^2+\tau (2G_M^V)^2}{1+\tau}\Delta
\right]\ ,
\end{eqnarray}
where

\begin{equation}
\Delta= \frac{\tau}{\kappa^ 2}\xi_F(1-\psi^2)
\left[\kappa\sqrt{1+\frac{1}{\tau}}+\frac{\xi_F}{3}(1-\psi^2)\right]
\end{equation}
and we have written the axial-vector response as  the sum of conserved (c.)
plus non conserved (n.c.) parts,
\begin{eqnarray}
\left(U_{CC}^A\right)_{\rm c.}
&=&
\frac{\kappa^2}{\tau}G_A^2\Delta
\\
\left(U_{CC}^A\right)_{\rm n.c.}
&=&
\frac{\lambda^2}{\tau}G_A'{}^2.
\end{eqnarray}
For $K=CL,LL$ we have
\begin{eqnarray}
U_{CL} &=& U_{CL}^V+\left(U_{CL}^A\right)_{\rm c.}
+\left(U_{CL}^A\right)_{\rm n.c.}
\\
U_{LL} &=& U_{LL}^V+\left(U_{LL}^A\right)_{\rm c.}
+\left(U_{LL}^A\right)_{\rm n.c.}\ ,
\end{eqnarray}
where the vector and conserved axial-vector parts are determined by
current conservation as
\begin{eqnarray}
U_{CL}^V &=& -\frac{\lambda}{\kappa}U_{CC}^V
\\
\left(U_{CL}^A\right)_{\rm c.}
&=& -\frac{\lambda}{\kappa}
\left(U_{CC}^A\right)_{\rm c.}
\\
U_{LL}^V &=& \frac{\lambda^2}{\kappa^2}U_{CC}^V
\\
\left(U_{LL}^A\right)_{\rm c.}
&=& \frac{\lambda^2}{\kappa^2}
\left(U_{CC}^A\right)_{\rm c.}\ ,
\end{eqnarray}
while the n.c. parts are
\begin{eqnarray}
\left(U_{CL}^A\right)_{\rm n.c.}
&=& -\frac{\lambda\kappa}{\tau}G_A'{}^2
\\
\left(U_{LL}^A\right)_{\rm n.c.}
&=& \frac{\kappa^2}{\tau}G_A'{}^2\ .
\end{eqnarray}
Finally the transverse responses are given by
\begin{eqnarray}
U_T &=& U_T^V+U_T^A
\\
U_T^V &=&  2\tau(2G_M^V)^2+\frac{(2G_E^V)^2+\tau (2G_M^V)^2}{1+\tau}\Delta
\\
U_T^A &=& 2(1+\tau)G_A^2 + G_A^2 \Delta
\\
U_{T'} &=& 2G_A(2G_M^V) \sqrt{\tau(1+\tau)}[1+\tilde{\Delta}]
\end{eqnarray}
with
\begin{equation}
\tilde{\Delta}=\sqrt{\frac{\tau}{1+\tau}}\frac{\xi_F(1-\psi^2)}{2\kappa}\ .
\end{equation}


\end{document}